\documentclass[prd,showpacs,showkeys,nofootinbib,twocolumn]{revtex4-1}

\usepackage{amsmath,amsfonts,amssymb}
\usepackage[colorlinks]{hyperref}
\usepackage{bm}% bold math

\def\subsvar{\delta_{(s)}}%Substantial variation

\begin{document}

\title{Invariant conserved currents in generalized gravity}

\author{Yuri N. Obukhov}
\email{obukhov@ibrae.ac.ru}
\affiliation{Theoretical Physics Laboratory, Nuclear Safety Institute, 
Russian Academy of Sciences, B.Tulskaya 52, 115191 Moscow, Russia} 

\author{Felipe Portales-Oliva}
\email{fportales@udec.cl}
\affiliation{Departamento de F\'isica, Universidad de Concepci\'on, 
Casilla 160-C, Concepci\'on, Chile}

\author{Dirk Puetzfeld}
\email{dirk.puetzfeld@zarm.uni-bremen.de}
\homepage{http://puetzfeld.org}
\affiliation{ZARM, University of Bremen, Am Fallturm, 28359 Bremen, Germany} 

\author{Guillermo F. Rubilar}
\email{grubilar@udec.cl}
\affiliation{Departamento de F\'isica, Universidad de Concepci\'on, 
Casilla 160-C, Concepci\'on, Chile }

\date{\today}

\begin{abstract}
We study conservation laws for gravity theories invariant under general coordinate transformations. The class of models under consideration includes Einstein's general relativity theory as a special case as well as its generalizations to non-Riemannian spacetime geometry and nonminimal coupling. We demonstrate that an arbitrary vector field on the spacetime manifold generates a current density that is conserved under certain conditions, and find the expression of the corresponding superpotential. For a family of models including nonminimal coupling between geometry and matter, we discuss in detail the differential conservation laws and the conserved quantities defined in terms of covariant multipole moments. We show that the equations of motion for the multipole moments of extended microstructured test bodies lead to conserved quantities that are closely related to the conserved currents derived in the field-theoretic framework.
\end{abstract}


%% 04.50.-h Higer-dimensional gravity and other theories of gravity 
%% 04.20.Fy Variational methods in general relativity
%% 04.20.Cv Fundamental problems and general formalism

\maketitle

\section{Introduction}

The correct understanding of the energy, momentum, and angular momentum in gravity theories is a prominent physical problem that has a long and complicated history. In order to solve this problem, numerous formalisms were developed to derive what are generally known as the conservation laws. There are two classes of conservation laws in any gravity theory. One class of conservation laws is formulated solely in terms of the dynamical variables that describe the gravitational field itself. These variables characterize the geometry of spacetime without involving additional structures of physical (non-geometrical) nature. The resulting conserved charges do not have tensor transformation properties under general coordinate transformations. 

There is, however, another class of conservation laws that lead to conserved charges and that turn out to be true scalars. In deriving such conservation laws, one usually has to deal with, besides the gravitational field variables, additional physical structures such as vector fields. As it is well known, vector fields generate diffeomorphisms on a spacetime manifold. What is more important, one can associate with a vector field $\zeta^i$ a current that is conserved under some conditions. In Einstein's general relativity theory this can be illustrated as follows. 

Given a covariantly conserved symmetric energy-momentum tensor, $\widetilde{\nabla}_iT_k{}^i = 0$, we find (contracting with $\zeta^k$) a relation: $\widetilde{\nabla}_iJ^i = {\frac 12}T^{ij}{\cal L}_\zeta g_{ij}$. Here we defined the current $J^i := \zeta^kT_k{}^i$; furthermore, $\widetilde{\nabla}_j$ is the Riemannian covariant derivative and ${\cal L}_\zeta$ is the Lie derivative along the vector field $\zeta^k$. If  $\zeta^k$ is a Killing vector, the current is conserved $\widetilde{\nabla}_iJ^i = 0$. This fact establishes a remarkable relation between the symmetries of the spacetime and the conserved currents generated by these symmetries. Physically, the vector field is usually related to the reference frame motion of an observer. 

In this paper we extend this observation to a general framework of gravity theories that encompass possible non-Riemannian geometries and nonminimal couplings of matter to the gravitational field. Theories of this kind attracted considerable attention in the literature.

The history of the construction of conserved quantities in relativistic gravity is rich and long. The earliest relevant construction is the Komar charge and its close relatives \cite{Komar:1959,Komar:1962,Wald:1994}. Essential contributions to this approach include \cite{Benn:1982,Julia:1998,Julia:2000,Fatibene:1999,Ferraris:1990,Ferraris:1994,Ferraris:2003,Katz:2006,Borowiec:1994,Chrusciel:1985,Petrov:2002,Deruelle:2005,Chang:1999,Chen:1999,Nester:2004,Barnich:2002}. Most of these works were confined to the purely Riemannian geometrical framework of Einstein's general relativity. Extensions to more general geometries were studied in \cite{Giachetta:1996,Sardanashvily:1997,Giachetta:1999,Mielke:2001,Hecht:1992,Aros1:2000,Aros2:2000} and more recently by us \cite{Obukhov:2006,Obukhov:2007,Obukhov:2008} and \cite{Petrov:2013,Lompay1:2013,Lompay2:2013}. 

The structure of the paper is as follows. In the next Sec.~\ref{pre} we give an overview of the basic geometrical notions and operations which underlie our study. Sec.~\ref{MAG} presents the essentials of generalized gravity theory. We give a detailed derivation of the currents associated with the vector fields (diffeomorphisms) on the spacetime manifold, and specify the conditions under which these currents are conserved. In particular, we demonstrate the importance of {\it generalized} Killing vector fields, the properties of which we discuss in Sec.~\ref{killing_sec}. In Sec.~\ref{MAG-nonminimal} we use our findings to obtain the generalized conserved current in metric-affine gravity (MAG) with a possible nonminimal coupling. Finally, we also show that the equations of motion for extended microstructured test bodies also admit a conserved quantity that is closely related (and has a similar structure) to the conserved currents derived in the field-theoretic framework. We conclude our paper in Sec.~\ref{conclusions_sec} with a discussion of the results obtained and with an outlook of their possible applications. 

Our notations and conventions are summarized at the end of the paper, in Appendix~\ref{notation}. 

\section{Preliminaries: spacetime geometry}\label{pre}

In the metric-affine theory of gravity, the gravitational physics is described by the dynamics of the geometrical structure of spacetime. The latter is encoded in two fields: the metric tensor $g_{ij}$ and an independent linear connection $\Gamma_{ki}{}^j$. The latter is not necessarily symmetric and/or compatible with the metric. From the geometrical point of view, the metric introduces lengths and angles of vectors, and thereby determines the distances (intervals) between points on the spacetime manifold. The connection introduces the notion of parallel transport and defines the covariant differentiation $\nabla_k$ of tensor fields. 

Locally, a spacetime diffeomorphism can be described as a small translation in the spacetime manifold, which technically is represented by the variation of the spacetime coordinates
\begin{equation}
\delta x^i = \epsilon\varepsilon^i(x).\label{dex}
\end{equation}
Here $\epsilon$ is an infinitesimal constant parameter and $\varepsilon^i(x)$ is an arbitrary, but finite, vector field. Under the action of diffeomorphisms, the geometrical variables transform as 
\begin{eqnarray}
\delta g_{ij} &=& -\,\epsilon(\partial_i\varepsilon^k)\,g_{kj} - \epsilon(\partial_j\varepsilon^k)\,g_{ik},\label{dgij}\\
\delta\Gamma_{ki}{}^j &=&  -\,\epsilon(\partial_k \varepsilon^l)\,\Gamma_{li}{}^j - \epsilon(\partial_i \varepsilon^l)\,\Gamma_{kl}{}^j\nonumber\\
&& +\,\epsilon(\partial_l \varepsilon^j)\,\Gamma_{ki}{}^l - \epsilon \partial^2_{ki}\varepsilon^j.\label{dG}
\end{eqnarray}

In general, the geometry of a metric-affine manifold is exhaustively characterized by three tensors: the curvature, the torsion and the nonmetricity. They are defined \cite{Schouten:1954} as follows:
\begin{eqnarray}
R_{kli}{}^j &:=& \partial_k\Gamma_{li}{}^j - \partial_l\Gamma_{ki}{}^j + \Gamma_{kn}{}^j \Gamma_{li}{}^n - \Gamma_{ln}{}^j\Gamma_{ki}{}^n,\label{curv}\\
T_{kl}{}^i &:=& \Gamma_{kl}{}^i - \Gamma_{lk}{}^i,\label{tors}\\ \label{nonmet}
Q_{kij} &:=& -\,\nabla_kg_{ij} = - \partial_kg_{ij} + \Gamma_{ki}{}^lg_{lj} + \Gamma_{kj}{}^lg_{il}.
\end{eqnarray}
The curvature and the torsion tensors determine the commutator of the covariant derivatives. For a tensor $A^{c_1 \dots c_k}{}_{d_1 \dots d_l}$ of arbitrary rank and index structure: 
\begin{eqnarray}
&& (\nabla_a\nabla_b - \nabla_b\nabla_a) A^{c_1 \dots c_k}{}_{d_1 \dots d_l} = -\,T_{ab}{}^e\nabla_e A^{c_1 \dots c_k}{}_{d_1 \dots d_l} \nonumber \\
&& + \sum^{k}_{i=1} R_{abe}{}^{c_i} A^{c_1 \dots e \dots c_k}{}_{d_1 \dots d_l} \nonumber \\
&& - \sum^{l}_{j=1}R_{abd_j}{}^{e} A^{c_1 \dots c_k}{}_{d_1 \dots e \dots d_l}. \label{commutator}
\end{eqnarray}
The Ricci tensor is introduced by $R_{ij} := R_{kij}{}^k$, and the curvature scalar is $R := g^{ij}R_{ij}$. 

A general metric-affine spacetime ($R_{kli}{}^j \neq 0$, $T_{kl}{}^i \neq 0$, $Q_{kij} \neq 0$) incorporates several other spacetimes as special cases.
The Riemannian connection $\widetilde{\Gamma}_{kj}{}^i$ is uniquely determined by the conditions of vanishing torsion and nonmetricity which yield explicitly 
\begin{equation}
\widetilde{\Gamma}_{kj}{}^i = {\frac 12}g^{il}(\partial_jg_{kl} + \partial_kg_{lj} - \partial_lg_{kj}).\label{Chr}
\end{equation}
Here and in the following, a tilde over a symbol denotes a Riemannian object (such as the curvature tensor) or a Riemannian operator (such as the covariant derivative) constructed from the Christoffel symbols (\ref{Chr}). The deviation of the geometry from the Riemannian one is then conveniently described by the {\it distortion} tensor 
\begin{equation}
N_{kj}{}^i := \widetilde{\Gamma}_{kj}{}^i - \Gamma_{kj}{}^i.\label{dist}
\end{equation}
The definitions (\ref{tors}) and (\ref{nonmet}) allows to find the distortion tensor in terms of the torsion and nonmetricity. Explicitly,
\begin{equation}
N_{kj}{}^i = -\,{\frac 12}(T_{kj}{}^i + T^i{}_{kj} + T^i{}_{jk})  + {\frac 12}(Q^i{}_{kj} - Q_{kj}{}^i - Q_{jk}{}^i).\label{NTQ}
\end{equation}
Conversely, one can use this to express the torsion and nonmetricity tensors in terms of the distortion,
\begin{eqnarray}
T_{kj}{}^i &=& -\,2N_{[kj]}{}^i,\label{TN}\\
Q_{kij} &=& -\,2N_{k(i}{}^lg_{j)l}.\label{QN}
\end{eqnarray}
Substituting (\ref{dist}) into (\ref{curv}), we find the relation between the non-Riemannian and the Riemannian curvature tensors,
\begin{equation}
R_{adc}{}^b = \widetilde{R}_{adc}{}^b - \widetilde{\nabla}_aN_{dc}{}^b + \widetilde{\nabla}_dN_{ac}{}^b + N_{an}{}^bN_{dc}{}^n - N_{dn}{}^bN_{ac}{}^n.\label{RRN}
\end{equation}
Applying the covariant derivative to (\ref{curv})-(\ref{nonmet}) and antisymmetrizing, we derive the Bianchi identities \cite{Schouten:1954}:
\begin{eqnarray}
\nabla_{[n}R_{kl]i}{}^j &=& T_{[kl}{}^m R_{n]mi}{}^j,\label{Dcurv}\\
\nabla_{[n}T_{kl]}{}^i &=& R_{[kln]}{}^i + T_{[kl}{}^m T_{n]m}{}^i,\label{Dtors}\\ 
\nabla_{[n}Q_{k]ij} &=& R_{nk(ij)}.\label{Dnonmet}
\end{eqnarray}

\subsection{Tensors, densities, and covariant differential operators} 

Along with tensors, an important role in physics is played by densities. A fundamental density $\sqrt{-g}$ is constructed from the determinant of the metric, $g=$det$g_{ij}$. Under diffeomorphisms (\ref{dex}) it transforms as
\begin{equation}
\delta\sqrt{-g} = - \epsilon\,(\partial_i\varepsilon^i)\,\sqrt{-g}.\label{ddet}
\end{equation}
This is a direct consequence of (\ref{dgij}). From any tensor $B_{i\dots}{}^{j\dots}$ one can construct a density ${\mathfrak B}_{i\dots}{}^{j\dots} = \sqrt{-g}B_{i\dots}{}^{j\dots}$. Although in this paper we will encounter only such objects, it is worthwhile to notice that not all densities are of this type, since they can have different weights. The fundamental density $\sqrt{-g}$ and all other densities discussed here have weight +1. See the exhaustive presentation in the book of Synge and Schild \cite{Synge:1978}. 

There are two kinds of covariant differential operators on the spacetime manifold, depending on whether the connection is involved or not. The Lie derivative ${\cal L}_\varepsilon$ is defined along any arbitrary vector field $\varepsilon^i$ and it maps tensors (densities) into tensors (densities) of the same rank. Let us recall the explicit form of the Lie derivative of the metric and the distortion:
\begin{eqnarray}
{\cal L}_\varepsilon g_{ij} &=& \varepsilon^k\partial_kg_{ij} + (\partial_i\varepsilon^k)g_{kj} + (\partial_j\varepsilon^k)g_{ik},\label{Lm}\\
{\cal L}_\varepsilon N_{kj}{}^i &=& \varepsilon^n\partial_nN_{kj}{}^i + (\partial_k\varepsilon^n)N_{nj}{}^i\nonumber\\ 
&& + (\partial_j\varepsilon^n)N_{kn}{}^i - (\partial_n\varepsilon^i)N_{kj}{}^n.\label{Ld}
\end{eqnarray}

In contrast, a covariant derivative $\nabla_k$ raises the rank of tensors (densities) and it is determined by the linear connection $\Gamma_{kj}{}^i$. Moreover, there are different covariant derivatives which arise for different connections that may coexist on the same manifold. 

A mathematical fact is helpful in this respect: every third rank tensor $X_{kj}{}^i$ defines a map of one connection into a different new connection 
\begin{equation}
\Gamma_{kj}{}^i\quad\longrightarrow\quad\Gamma_{kj}{}^i + X_{kj}{}^i.\label{G2G}
\end{equation} 
There are important special cases of such a map. One example is obtained for $X_{kj}{}^i = N_{kj}{}^i$: then the connection $\Gamma_{kj}{}^i$ is mapped into the Riemannian Christoffel symbols, $\widetilde{\Gamma}_{kj}{}^i = \Gamma_{kj}{}^i + N_{kj}{}^i$, in accordance with (\ref{dist}). 

Another interesting case arises for $X_{kj}{}^i = T_{jk}{}^i$. The result of the mapping
\begin{equation}
\overline{\Gamma}_{kj}{}^i = \Gamma_{kj}{}^i + T_{jk}{}^i = \Gamma_{kj}{}^i + \Gamma_{jk}{}^i - \Gamma_{kj}{}^i = \Gamma_{jk}{}^i \label{tr}
\end{equation}
is then called a {\it transposed connection}, or associated connection, see \cite{Lichnerowicz:1977,Trautman:1973}. 

The importance of the transposed connection is manifest in the following observation. Although the Lie derivative is a covariant operator -- this is not apparent since it is based on partial derivatives -- one can make everything explicitly covariant by noticing that it is possible to recast (\ref{Lm}) and (\ref{Ld}) into equivalent forms
\begin{eqnarray}
{\cal L}_\varepsilon g_{ij} &=& \varepsilon^k\nabla_kg_{ij} + (\overline{\nabla}_i\varepsilon^k)g_{kj} + (\overline{\nabla}_j\varepsilon^k)g_{ik},\label{LmC}\\
{\cal L}_\varepsilon N_{kj}{}^i &=& \varepsilon^n\nabla_nN_{kj}{}^i + (\overline{\nabla}_k\varepsilon^n)N_{nj}{}^i\nonumber\\
&& + (\overline{\nabla}_j\varepsilon^n)N_{kn}{}^i - (\overline{\nabla}_n\varepsilon^i)N_{kj}{}^n.\label{LdC}
\end{eqnarray}
By the same token we can ``covariantize'' the Lie derivatives for all other tensors of any structure and of arbitrary rank. 

A more nontrivial (and less known) fact is that we can define the Lie derivatives also for objects which are not tensors. In particular, the Lie derivative of the connection then reads \cite{Lichnerowicz:1977}:
\begin{eqnarray}
{\cal L}_\varepsilon\Gamma_{kj}{}^i &=& \varepsilon^l\partial_l\Gamma_{kj}{}^i + (\partial_k\varepsilon^l)\,\Gamma_{lj}{}^i + (\partial_j\varepsilon^l)\,\Gamma_{kl}{}^i  \nonumber \\
&& - \,(\partial_l\varepsilon^i)\,\Gamma_{kj}{}^l	 + \partial^2_{kj}\varepsilon^i  \\
&=& \nabla_k\overline{\nabla}_j\varepsilon^i - R_{klj}{}^i\varepsilon^l.\label{LieGam}
\end{eqnarray}
This quantity measures the noncommutativity of the Lie derivative with the covariant derivative
\begin{eqnarray}
&&({\cal L}_\varepsilon \nabla_k - \nabla_k{\cal L}_\varepsilon) A^{c_1 \dots c_k}{}_{d_1 \dots d_l} \nonumber\\
&& = \,\sum^{k}_{i=1} ({\cal L}_\varepsilon\Gamma_{kb}{}^{c_i}) A^{c_1 \dots b \dots c_k}{}_{d_1 \dots d_l} 
\nonumber\\
&& - \sum^{l}_{j-1} ({\cal L}_\varepsilon\Gamma_{kd_j}{}^{b}) A^{c_1 \dots c_k}{}_{d_1 \dots b \dots d_l}. \label{commLD}
\end{eqnarray}

The connection $\Gamma_{kj}{}^i$, the transposed connection  $\overline{\Gamma}_{kj}{}^i$, and the Riemannian connection  $\widetilde{\Gamma}_{kj}{}^i$ define the three respective covariant derivatives: $\nabla_k$, $\overline{\nabla}_k$, and $\widetilde{\nabla}_k$. 

We will assume that these differential operators act on tensors. In addition, we will need the covariant operators that act on densities. For an arbitrary tensor density ${\mathfrak B}^n{}_{i\dots}{}^{j\dots}$ we introduce the covariant divergence 
\begin{equation}
\widehat{\nabla}{}_n{\mathfrak B}^n{}_{i\dots}{}^{j\dots} := \partial_n{\mathfrak B}^n{}_{i\dots}{}^{j\dots} + \Gamma_{nl}{}^j{\mathfrak B}^n{}_{i\dots}{}^{l\dots} - \Gamma_{ni}{}^l {\mathfrak B}^n{}_{l\dots}{}^{j\dots},\label{dA}
\end{equation}
which produces again a tensor density. We denote a similar differential operation constructed with the help of the Riemannian connection by
\begin{equation}
\check{\nabla}{}_n{\mathfrak B}^n{}_{i\dots}{}^{j\dots}:= \partial_n{\mathfrak B}^n{}_{i\dots}{}^{j\dots} + \widetilde{\Gamma}_{nl}{}^j{\mathfrak B}^n{}_{i\dots}{}^{l\dots} - \widetilde{\Gamma}_{ni}{}^l{\mathfrak B}^n{}_{l\dots}{}^{j\dots},\label{dAc}
\end{equation}
When ${\mathfrak B}^n{}_{i\dots}{}^{j\dots} = \sqrt{-g}B^n{}_{i\dots}{}^{j\dots}$, we find 
\begin{eqnarray}
\widehat{\nabla}_n{\mathfrak B}^n{}_{i\dots}{}^{j\dots} &=& \sqrt{-g}\,{\stackrel * \nabla}{}_i B^n{}_{i\dots}{}^{j\dots}, \label{nablastar}\\
\check{\nabla}_n{\mathfrak B}^n{}_{i\dots}{}^{j\dots} &=& \sqrt{-g}\,\widetilde{\nabla}{}_nB^n{}_{i\dots}{}^{j\dots}, \label{nablatilde}
\end{eqnarray}
where we introduced a modified covariant derivative
\begin{equation}
{\stackrel * \nabla}{}_i := \nabla_i + N_{ki}{}^k.\label{dstar}
\end{equation}
A remark explaining our notation is in order. As one knows, a tensor is special case of a density with zero weight. So, strictly speaking, the introduction of the numerous superscript accents above may be viewed as something redundant. However, we find it more convenient to explicitly distinguish densities from tensors with the help of the ``Fraktur'' font for the former, and the ``Roman'' font for the latter. Accordingly we use different accented symbols for the derivatives. In particular, one should note that (\ref{dstar}) acts on tensors, in contrast to (\ref{dA}) and (\ref{dAc}) which act on densities.

\subsection{Matter variables}

We will not specialize the discussion of matter to any particular physical field. It will be more convenient to describe matter by a generalized field $\psi^A$. The range of the indices $A,B,\dots$ is not important in our study. However, we do need to know the behavior of the matter field under spacetime diffeomorphisms. We assume that under the transformation (\ref{dex}), these fields satisfy
\begin{equation}
\delta\psi^A = -\,\epsilon(\partial_i\varepsilon^j)\,(\sigma^A{}_B)_j{}^i\,\psi^B.\label{dpsiA}
\end{equation}
Here $(\sigma^A{}_B)_j{}^i$ are the generators of general coordinate transformations that satisfy the commutation relations
\begin{eqnarray}\nonumber
&& (\sigma^A{}_C)_j{}^i(\sigma^C{}_B)_l{}^k - (\sigma^A{}_C)_l{}^k (\sigma^C{}_B)_j{}^i\\
&& = (\sigma^A{}_B)_l{}^i\,\delta^k_j - (\sigma^A{}_B)_j{}^k \,\delta^i_l.\label{comms}
\end{eqnarray}
We immediately recognize in (\ref{comms}) the Lie algebra of the general linear group $GL(4,R)$. This fact is closely related to the standard gauge-theoretic interpretation \cite{Hehl:1995} of metric-affine gravity as the gauge theory of the general affine group $GA(4,R)$, which is a semidirect product of spacetime translation group times $GL(4,R)$.

The transformation properties (\ref{dpsiA}) determine the form of the covariant and the Lie derivative of a matter field: 
\begin{eqnarray}\label{Dpsi}
\nabla_k\psi^A &:=& \partial_k\psi^A -\Gamma_{ki}{}^j\,(\sigma^A{}_B)_j{}^i\,\psi^B,\\ \label{Lpsi}
{\cal L}_\varepsilon\psi^A &:=& \varepsilon^k\partial_k\psi^A + (\partial_i\varepsilon^j)(\sigma^A{}_B)_j{}^i\,\psi^B \\
&=& \varepsilon^k\nabla_k\psi^A + (\overline{\nabla}_i\varepsilon^j)(\sigma^A{}_B)_j{}^i\,\psi^B.\label{Lp}
\end{eqnarray}
The commutators of these differential operators read
\begin{eqnarray}
(\nabla_k\nabla_l - \nabla_l\nabla_k)\psi^A &=& -\,R_{klj}{}^i(\sigma^A{}_B)_i{}^j\psi^B\nonumber\\
&& - T_{kl}{}^i\nabla_i\psi^A,\label{comDDpsi}\\
 ({\cal L}_\varepsilon \nabla_k - \nabla_k{\cal L}_\varepsilon)\psi^A &=& -\,({\cal L}_\varepsilon\Gamma_{kj}{}^i)(\sigma^A{}_B)_i{}^j\psi^B.\label{comDLpsi}
\end{eqnarray}

\section{Metric-affine gravity: field equations and currents}\label{MAG}

The explicit form of the dynamical equations of the gravitational field is irrelevant for the conservation laws that will form the basis for the derivation of the test body equations of motion. However, for completeness, we discuss here the field equations of a general metric-affine theory of gravity. The standard understanding of MAG is its interpretation as a gauge theory based on the general affine group $GA(4,R)$, which is a semidirect product of the general linear group $GL(4,R)$, and the group of local translations \cite{Hehl:1995}. The corresponding gauge-theoretic formalism generalizes the approach of Sciama and Kibble \cite{Sciama:1962,Kibble:1961}; for more details about gauge gravity theories, see \cite{Gronwald:1997,ObukhovS:2006,Blagojevic:2002,Hehl:2013}. Besides its many interesting properties, MAG offers the possibility of a unification of gravity with other physical interactions on the same gauge-theoretic principles, and contributes to the solution of the quantum gravity quest with encouraging attempts to construct a renormalizable theory of the quantized gravitational field \cite{Dell:1986,Lee:1990,Lee:1992,Pagani:2014,Pagani:2015}. 

In the standard formulation of MAG as a gauge theory \cite{Hehl:1995}, the gravitational gauge potentials are identified with the metric, coframe, and the linear connection. The corresponding gravitational field strengths are then the nonmetricity, the torsion, and the curvature, respectively. 

It will be convenient to describe all the dynamical variables -- including the gravitational (geometrical) and material fields -- collectively by means of a multiplet, which we denote by $\Phi^J = (g_{ij},\Gamma_{ki}{}^j,\psi^A)$. The range of the multi-index $J$ is not important at present and will be specified when needed.

\subsection{Lagrange-Noether analysis}

The dynamics of the interacting gravitational and matter fields is determined by a general action 
\begin{equation}
I = \int\,d^4x\,{\mathfrak L}(\Phi^J,\partial_i\Phi^J).\label{action}
\end{equation}
The Lagrangian density ${\mathfrak L}$ depends arbitrarily on its arguments. 
Let us investigate the variation of the action under transformations of the spacetime coordinates and the fields which, quite generally, read as follows:
\begin{eqnarray}
x'^i (x) &=& x^i + \delta x^i,\label{dx}\\
\Phi'^J(x') &=& \Phi^J(x) + \delta\Phi^J(x).\label{dP}
\end{eqnarray} 
Then, using the substantial variation defined by $\subsvar\Phi^J := \Phi'^J(x) - \Phi^J(x) = \delta\Phi^J - \delta x^k\partial_k\Phi^J$, we derive in a standard way the total variation of the action
\begin{eqnarray}
\delta I &=& \int d^4x \left[\subsvar{\mathfrak L} + \partial_i\left({\mathfrak L}\delta x^i\right)\right]. \label{masterV}
\end{eqnarray}
Assuming the invariance of the action under (\ref{dx}) and (\ref{dP}), we find the so-called Lie differential equation
\begin{equation}
\subsvar{\mathfrak L} + \partial_i\left({\mathfrak L}\delta x^i\right) = 0.\label{lieDE}
\end{equation}
With the help of the chain rule we can write
\begin{equation}
\subsvar{\mathfrak L} = {\frac {\partial {\mathfrak L}}{\partial\Phi^J}}\subsvar\Phi^J + {\frac {\partial {\mathfrak L}}{\partial(\partial_i\Phi^J)}}\subsvar\partial_i\Phi^J,\label{chain}
\end{equation}
and using the commutativity of the substantial variation with the partial derivative, $\subsvar\partial_i = \partial_i\subsvar$, we recast (\ref{lieDE}) into a balance equation
\begin{equation}
{\frac {\delta {\mathfrak L}}{\delta\Phi^J}} \,\subsvar\Phi^J + \partial_i\left({\mathfrak L}\,\delta x^i + {\frac {\partial {\mathfrak L}}{\partial(\partial_i\Phi^J)}}\,\subsvar \Phi^J \right) = 0.\label{balance}
\end{equation}
Here we denote the variational derivative
\begin{equation}\label{var}
{\frac {\delta {\mathfrak L}}{\delta\Phi^J}} := {\frac {\partial {\mathfrak L}}{\partial\Phi^J}} - \partial_i\left({\frac {\partial {\mathfrak L}}{\partial(\partial_i\Phi^J)}}\right),
\end{equation}
as usual. When the variational derivatives are put equal to zero, we obtain the system of classical field equations. The configurations of variables that satisfy ${\delta {\mathfrak L}}/{\delta\Phi^J} = 0$ are called ``on-shell''.

The balance equation (\ref{balance}) is an identity, i.e.\ it is valid for all configurations of the gravitational and matter fields irrespectively of the fact that they are ``on-shell'' or ``off-shell''. Equation (\ref{balance}) gives rise to various identities and conservation laws which we derive in this paper. 

We can apply (\ref{balance}) to different symmetries of the physical system under consideration. Of particular importance is the diffeomorphism (or general coordinate) invariance of the action (\ref{action}). Substituting (\ref{dex}), (\ref{dgij}), (\ref{dG}), and (\ref{dpsiA}) into (\ref{balance}), we recast the latter (after dropping the overall infinitesimal constant $\epsilon$) into
\begin{equation}
\Omega_k\,\varepsilon^k + \Omega_k{}^i\,\partial_i \varepsilon^k + {\frac 12}\Omega_k{}^{ij}\,\partial^2_{ij}\varepsilon^k + {\frac 13}\Omega_k{}^{ijl}\,\partial^3_{ijl}\varepsilon^k = 0.\label{Omegas0}
\end{equation}
The functions $\Omega_k{}^{i_1\dots i_n}$ (with $n=0,1,2,3$) are determined by the Lagrangian of the theory. Their explicit form is given in Appendix \ref{A1}.

In view of the arbitrariness of the vector field $\varepsilon^k$ and their derivatives, we obtain a set of Noether identities
\begin{equation}
\Omega_k{}^{i_1\dots i_n} = 0\label{Noetherall}
\end{equation}
for $n=0,1,2,3$. 

\subsection{Generalized current}

Every vector field on the spacetime manifold induces a current which is conserved under certain conditions. We can derive it with the help of the balance equation (\ref{balance}) as follows. Let us consider a map of the manifold (diffeomorphism) induced by a vector field, as in Eq. \eqref{dex}. Locally, such a diffeomorphism acts on the gravitational and matter variables $\Phi^J = (g_{ij},\Gamma_{ki}{}^j,\psi^A)$ by means of the Lie derivatives so that
\begin{eqnarray}
\subsvar g_{ij} &=& -\epsilon{\cal L}_\varepsilon g_{ij},\label{Zg}\\ 
\subsvar \Gamma_{kj}{}^i &=& -\epsilon{\cal L}_\varepsilon \Gamma_{kj}{}^i,\label{GZ}\\ 
\subsvar \psi^A &=& -\epsilon{\cal L}_\varepsilon \psi^A.\label{pZ} 
\end{eqnarray}
Inserting this into (\ref{balance}) and dividing by the infinitesimal parameter $\epsilon$, we can define the current density
\begin{eqnarray}
{\mathfrak J}^i &:=&   {\frac {\partial {\mathfrak L}}{\partial \partial_ig_{kl}}}{\cal L}_\varepsilon g_{kl} +\,{\frac {\partial {\mathfrak L}}{\partial \partial_i\Gamma_{kn}{}^m}}{\cal L}_\varepsilon\Gamma_{kn}{}^m \nonumber\\
&&  + {\frac {\partial {\mathfrak L}}{\partial \partial_i\psi^A}}{\cal L}_\varepsilon \psi^A  - {\mathfrak L}\,\varepsilon^i.\label{Jdef}
\end{eqnarray}
and observe that it satisfies
\begin{eqnarray}
\partial_i{\mathfrak J}^i = -{\frac {\delta {\mathfrak L}}{\delta g_{kl}}}{\cal L}_\varepsilon g_{kl} - {\frac {\delta {\mathfrak L}}{\delta\Gamma_{kn}{}^m}}{\cal L}_\varepsilon\Gamma_{kn}{}^m - {\frac {\delta {\mathfrak L}}{\delta\psi^A}}{\cal L}_\varepsilon \psi^A. \nonumber \\ \label{Jbalance}
\end{eqnarray}
Using (\ref{Lm}), (\ref{LieGam}), and (\ref{Lp}), we can display the structure of the current as follows 
\begin{equation}
{\mathfrak J}^i = \varepsilon^m{\mathfrak J}_m{}^i + (\overline{\nabla}_n\varepsilon^m){\mathfrak J}_m{}^{ni} + (\nabla_k\overline{\nabla}_n\varepsilon^m){\mathfrak J}_m{}^{kni}.\label{J1}
\end{equation}
Here we introduced the densities
\begin{eqnarray}\label{J2}
{\mathfrak J}_m{}^i &:=& {\frac {\partial {\mathfrak L}}{\partial \partial_i\psi^A}}\,\nabla_m\psi^A - {\mathfrak L}\,\delta^i_m\nonumber\\
&& -\,{\frac {\partial {\mathfrak L}}{\partial \partial_ig_{kl}}}\,Q_{mkl} - {\frac {\partial {\mathfrak L}}{\partial \partial_i\Gamma_{kl}{}^n}}\,R_{kml}{}^n,\\
{\mathfrak J}_m{}^{ni} &:=& 2{\frac {\partial {\mathfrak L}}{\partial \partial_ig_{ln}}}\,g_{lm} +  {\frac {\partial {\mathfrak L}}{\partial \partial_i\psi^A}}\,(\sigma^A{}_B)_m{}^n\psi^B,\label{J3}\\
{\mathfrak J}_m{}^{kni} &:=& {\frac {\partial {\mathfrak L}}{\partial \partial_i\Gamma_{kn}{}^m}}.\label{J4}
\end{eqnarray}

\subsection{Current and superpotential}

Expanding the Lie derivatives in (\ref{Jdef}), we can recast the current into a different equivalent form. Namely, a lengthy direct computation yields
\begin{eqnarray}
{\mathfrak J}^i &=& \Omega_k{}^i\varepsilon^k + \Omega_k{}^{ni}\partial_n\varepsilon^k + \Omega_k{}^{mni}\partial^2_{mn}\varepsilon^k\nonumber\\
&& - \,\varepsilon^j\left[2{\frac {\delta {\mathfrak L}}{\delta g_{ik}}}g_{jk} + {\frac {\delta {\mathfrak L}}{\delta \psi^A}}(\sigma^A{}_B)_j{}^i\psi^B - \widehat{\nabla}_k{\frac {\delta {\mathfrak L}}{\delta \Gamma_{ki}{}^j}}\right]\nonumber\\
&& - \,(\overline{\nabla}_m\varepsilon^n){\frac {\delta {\mathfrak L}}{\delta \Gamma_{im}{}^n}} + \partial_j{\mathfrak K}^{ij}.\label{Ji}
\end{eqnarray}
Here we introduced a new density
\begin{eqnarray}
{\mathfrak K}^{ij} &:=& -\,\varepsilon^k\left[2{\frac {\partial {\mathfrak L}}{\partial\partial_jg_{il}}}g_{lk} + {\frac {\partial {\mathfrak L}}{\partial \partial_j\psi^A}}(\sigma^A{}_B)_k{}^i\psi^B \right.\nonumber\\
&& + \,{\frac {\partial {\mathfrak L}}{\partial \Gamma_{ji}{}^k}} + {\frac {\partial {\mathfrak L}}{\partial\partial_j\Gamma_{in}{}^m}}\Gamma_{kn}{}^m + {\frac {\partial {\mathfrak L}}{\partial\partial_j\Gamma_{ni}{}^m}}\Gamma_{nk}{}^m\nonumber\\
&&\left. - \,{\frac {\partial {\mathfrak L}}{\partial\partial_j\Gamma_{nm}{}^k}}\Gamma_{nm}{}^i - \partial_l{\frac {\partial {\mathfrak L}}{\partial\partial_{(l}\Gamma_{j)i}{}^k}}\right]\nonumber\\ 
&& - \,(\partial_n\varepsilon^k)\left({\frac {\partial{\mathfrak L}}{\partial\partial_j\Gamma_{in}{}^k}} + {\frac {\partial{\mathfrak L}}{\partial\partial_{(n}\Gamma_{j)i}{}^k}}\right).\label{Kij}
\end{eqnarray}

The first line on the right-hand side of (\ref{Ji}) vanishes for the diffeomorphism invariant (generally covariant) theories in view of the Noether identities (\ref{Noetherall}). As a result, we obtain ``on-shell'' (i.e., when $\delta {\mathfrak L}/\delta g_{ij} = 0$, $\delta {\mathfrak L}/\delta \Gamma_{kj}{}^i = 0$, and $\delta {\mathfrak L}/\delta \psi^A = 0$): 
\begin{equation}
{\mathfrak J}^i = \partial_j{\mathfrak K}^{ij}.\label{JK}
\end{equation}
This means that on the classical field equations, the current ${\mathfrak J}^i$ is derived from the {\it superpotential} density ${\mathfrak K}^{ij}$. 

\subsection{Explicitly covariant current}

It is not obvious that the complicated expressions (\ref{Jdef}), (\ref{Ji}), and (\ref{Kij}) are truly covariant objects under the action of diffeomorphisms. A direct demonstration is possible, but it is somewhat long to present it here. Instead, we will show the covariance by specifying the form of the Lagrangian. Intuitively it is clear (and can be rigorously proven) that the Lagrangian of a generally covariant theory should be a function of covariant objects. This means, in particular, that the derivatives of the basic variables  $\Phi^J = (g_{ij},\Gamma_{ki}{}^j,\psi^A)$ should only appear in the form of explicitly covariant combinations. In simple terms,
\begin{equation}
{\mathfrak L}(\Phi^J,\partial\Phi^J) = {\mathfrak L}(g_{ij}, Q_{kij}, T_{ij}{}^k, R_{ijk}{}^l, \psi^A, \nabla_i\psi^A).\label{Lcov}
\end{equation}
We then immediately compute the partial derivatives
\begin{eqnarray}
{\frac {\partial {\mathfrak L}}{\partial \partial_jg_{il}}} &=& -\,{\frac {\partial {\mathfrak L}}{\partial Q_{jil}}},\label{dL1}\\
{\frac {\partial {\mathfrak L}}{\partial \partial_j\Gamma_{in}{}^m}} &=& 2\,{\frac {\partial {\mathfrak L}}{\partial R_{jin}{}^m}},\label{dL2}\\
{\frac {\partial {\mathfrak L}}{\partial \partial_j\psi^A}} &=& {\frac {\partial {\mathfrak L}}{\partial \nabla_j\psi^A}},\label{dL3}\\
{\frac {\partial {\mathfrak L}}{\partial \Gamma_{ki}{}^j}} &=& 2{\frac {\partial {\mathfrak L}}{\partial Q_{kil}}}g_{jl} + 2{\frac {\partial {\mathfrak L}}{\partial T_{ki}{}^j}} - {\frac {\partial {\mathfrak L}}{\partial \nabla_k\psi^A}}(\sigma^A{}_B)_j{}^i\psi^B\nonumber\\
&& + 2{\frac {\partial {\mathfrak L}}{\partial R_{kmn}{}^j}}\Gamma_{mn}{}^i + 2{\frac {\partial {\mathfrak L}}{\partial R_{mki}{}^n}}\Gamma_{mj}{}^n.\label{dL4}
\end{eqnarray}
Substituting these expressions, we obtain the explicitly covariant current 
\begin{eqnarray}
{\mathfrak J}^i &=& -\varepsilon^j\left[2{\frac {\delta {\mathfrak L}}{\delta g_{ik}}}g_{jk} + {\frac {\delta {\mathfrak L}}{\delta \psi^A}}(\sigma^A{}_B)_j{}^i\psi^B - \widehat{\nabla}_k{\frac {\delta {\mathfrak L}}{\delta \Gamma_{ki}{}^j}}\right]\nonumber\\
&& - \,(\overline{\nabla}_m\varepsilon^n){\frac {\delta {\mathfrak L}}{\delta \Gamma_{im}{}^n}} + \check{\nabla}_j{\mathfrak K}^{ij}.\label{Jcov}
\end{eqnarray}
and the explicitly covariant superpotential
\begin{equation}\label{Kcov}
{\mathfrak K}^{ij} = 2\,\varepsilon^k\,{\frac{\partial {\mathfrak L}}{\partial T_{ij}{}^k}} + 2\,(\overline{\nabla}_n\varepsilon^k)\,{\frac {\partial{\mathfrak L}}{\partial R_{ijn}{}^k}}.
\end{equation}

 In our work \cite{Obukhov:2006}, we studied the definition of conserved currents within models with local Lorentz invariance and vanishing nonmetricity. In those models, one can define different conserved currents depending on the choice of a so-called generalized Lie derivative. See \cite{Obukhov:2006,Obukhov:2008} for further details. On the other hand, the current \eqref{Jcov} and the superpotential \eqref{Kcov} are defined for any metric-affine model. Comparing both results for the restricted case of vanishing nonmetricity, one can verify after some straightforward algebra, that the superpotential \eqref{Kcov} reduces to the components of the potential 2-form $H$ of \cite{Obukhov:2006}, provided one uses the Yano choice for the generalized Lie derivative \cite{Yano:1955,Kosmann1:1966,Kosmann2:1966}. Therefore, the corresponding currents coincide when the classical field equations for the total system are satisfied. As a consequence, in the case of the Hilbert-Einstein Lagrangian, the conserved quantities reduce, in vacuum, to the well known Komar charges \cite{Komar:1959,Komar:1962}.

\subsection{Conservation of the general current}

The current ${\mathfrak J}^i$ defined in \eqref{Jdef}, or equivalently in \eqref{Jcov}, is conserved when the right-hand side of the balance equation \eqref{Jbalance} vanishes,
\begin{equation}
\partial_i{\mathfrak J}^i = 0.\label{consJ}
\end{equation} 
This is the case, for instance, for the ``on-shell'' field configuration when both the gravitational and matter variables satisfy the classical field equations $\delta {\mathfrak L}/\delta g_{ij} = 0$, $\delta {\mathfrak L}/\delta \Gamma_{kj}{}^i = 0$, and $\delta {\mathfrak L}/\delta \psi^A = 0$. Alternatively, the current is conserved, even ``off-shell'', when the Lie derivatives of all field variables vanish along a particular choice of the vector field $\varepsilon^i$. In these cases, when the vector field is not arbitrary, but some particular special case, we denote it by $\zeta^i$, so that it satisfies ${\cal L}_\zeta g_{ij} = 0$, ${\cal L}_\zeta\Gamma_{kj}{}^i = 0$, and ${\cal L}_\zeta \psi^A = 0$. 

The last assumptions are, however, a bit too strong. A milder condition for the conservation of the current is the ``on-shell''  ansatz for the matter fields, $\delta {\mathfrak L}/\delta \psi^A = 0$, combined with the vanishing of only the Lie derivatives for the gravitational fields along a vector field $\zeta^i$: ${\cal L}_\zeta g_{ij} = 0$ and ${\cal L}_\zeta\Gamma_{kj}{}^i = 0$. We discuss the geometrical meaning of the latter conditions in Sec. \ref{killing_sec}.

Note also that \eqref{consJ} can be written in terms of the covariant derivatives $\widehat{\nabla}$ and $\check{\nabla}$ since, by the general definitions \eqref{dA} and \eqref{dAc}, we have
\begin{equation}
\widehat{\nabla}_i{\mathfrak J}^{i}=\check{\nabla}_i{\mathfrak J}^{i} =\partial_i{\mathfrak J}^i .\label{consJcov}
\end{equation}

\section{Symmetries in MAG: generalized Killing vectors}\label{killing_sec}

As is well known, symmetries of a Riemannian spacetime are generated by Killing vector fields. Each such field defines a so-called {\it motion} of the spacetime manifold, that is a diffeomorphism which preserves the metric $g_{ij}$. 

Suppose $\zeta^i$ is a Killing vector field. By definition, it satisfies
\begin{equation}
\widetilde{\nabla}_i\zeta_j + \widetilde{\nabla}_j\zeta_i = 0.\label{K1}
\end{equation}
By differentiation, we derive from this the second covariant derivative 
\begin{equation}
\widetilde{\nabla}_i\widetilde{\nabla}_j\zeta_k = \widetilde{R}_{jki}{}^l\zeta_l.\label{K2}
\end{equation}
We apply another covariant derivative and antisymmetrize: 
\begin{equation}
\widetilde{\nabla}_{[n}\widetilde{\nabla}_{i]}\widetilde{\nabla}_j\zeta_k = \widetilde{\nabla}_{[n}(\widetilde{R}_{|jk|i]}{}^l\zeta_l).\label{K2a}
\end{equation}
After some algebra, the last equation is recast into
\begin{eqnarray}
\zeta^n\widetilde{\nabla}_n\widetilde{R}_{ijkl} + \widetilde{R}_{njkl}\widetilde{\nabla}_i\zeta^n + \widetilde{R}_{inkl}\widetilde{\nabla}_j\zeta^n\nonumber\\ + \widetilde{R}_{ijnl}\widetilde{\nabla}_k\zeta^n + \widetilde{R}_{ijkn}\widetilde{\nabla}_l\zeta^n = 0.\label{K3}
\end{eqnarray}

Equations (\ref{K1}), (\ref{K2}), and (\ref{K3}) have a geometrical meaning:
\begin{eqnarray}
{\cal L}_\zeta g_{ij} &=& 0,\label{Lg}\\
{\cal L}_\zeta \widetilde{\Gamma}_{ij}{}^k &=& 0,\label{LG}\\
{\cal L}_\zeta \widetilde{R}_{ijkl} &=& 0.\label{LR}
\end{eqnarray}
That is, the Lie derivatives along the Killing vector field $\zeta$ vanish for all Riemannian geometrical objects. Moreover, one can show that the same is true for all higher covariant derivatives of the Riemannian curvature tensor \cite{Yano:1955}:
\begin{equation}
{\cal L}_\zeta \left(\widetilde{\nabla}_{n_1}\dots \widetilde{\nabla}_{n_N}\widetilde{R}_{ijkl}\right) = 0.\label{LDR} 
\end{equation}

It is worthwhile to mention that in the Riemannian framework of Einstein's general relativity, one can define various symmetries generated by the vector fields (diffeomorphisms). For example, if $\zeta$ does not satisfy (\ref{Lg}) but fulfills (\ref{LG}), such a vector field is not a Killing but a so-called called {\it affine collineation}. Alternatively, if both (\ref{Lg}) and (\ref{LG}) are not true but $\zeta$ is characterized by the property (\ref{LR}), it is called a {\it curvature collineation}. Along with the standard Killing vector fields, these new fields contain important information about the symmetries of the spacetime, and it is possible to use them to define additional conservation laws. A comprehensive discussion of such symmetries (and related ones such as homothetic, conformal, projective, Ricci collineations etc) can be found in \cite{Katzin:1969,Hall:2004}, and different applications are studied in \cite{Collinson:1970,Hojman:1986,Harte:2008}, for example.

Let us generalize the notion of a symmetry to the metric-affine spacetime. We take an ordinary Killing vector field $\zeta$ and postulate the vanishing of the Lie derivative 
\begin{equation}
{\cal L}_\zeta N_{kj}{}^i = 0\label{LieN}
\end{equation}
of the distortion tensor. Combining this with (\ref{NTQ}) and (\ref{LG}), we find an equivalent formulation
\begin{eqnarray}
{\cal L}_\zeta g_{ij} &=& 0,\label{LgMAG}\\
{\cal L}_\zeta \Gamma_{ij}{}^k &=& 0.\label{LGMAG}
\end{eqnarray}
We call a vector field that satisfies (\ref{LgMAG}) and (\ref{LGMAG}) a {\it generalized Killing vector} of the metric-affine spacetime. By definition, such a $\zeta$ generates a diffeomorphism of the spacetime manifold that is simultaneously an isometry (\ref{LgMAG}) and an isoparallelism (\ref{LGMAG}). 

Since the Lie derivative along a Killing vector commutes with the covariant derivative, ${\cal L}_\zeta\widetilde{\nabla}_i = \widetilde{\nabla}_i {\cal L}_\zeta$, see (\ref{commLD}), we conclude from (\ref{RRN}) and (\ref{LR}) that the generalized Killing vector leaves the non-Riemannian curvature tensor invariant
\begin{equation}
{\cal L}_\zeta R_{klj}{}^i = 0.\label{LRMAG}
\end{equation}
It is also straightforward to verify that 
\begin{equation}
{\cal L}_\zeta \left(\nabla_{n_1}\dots \nabla_{n_N}R_{klj}{}^i\right) = 0\label{LDRMAG} 
\end{equation}
for any number of covariant derivatives of the curvature.

Finally, combining (\ref{TN}) and (\ref{QN}) with (\ref{Lg}) and (\ref{LieN}), we verify that
\begin{eqnarray}
{\cal L}_\zeta T_{kj}{}^i &=& 0,\label{LTMAG}\\
{\cal L}_\zeta Q_{kij} &=& 0.\label{LQMAG}
\end{eqnarray}
Summarizing, the Lie derivative of {\it all} main geometrical objects vanishes along a generalized Killing vector field. 

Later we will show that generalized Killing vectors have an important property: they induce conserved quantities on the metric-affine spacetime. 

\section{Modified MAG models with a possible non-minimal coupling} \label{MAG-nonminimal}

\subsection{Gravitational field dynamics}

The general formalism which we developed in Sec.~\ref{MAG} will now be applied to specific models of the gravitational field coupled with matter. However, before we go into technical details, the following remark is in order. Strictly speaking, the metric-affine framework is applicable not only to generalized gravity theories which are based on the non-Riemannian spacetime geometries, but it is equally useful for the study of Einstein's GR and its modifications in a purely Riemannian geometrical context. Technically, this requires the introduction of additional variables which play the role of Lagrange multipliers, and this changes the physical meaning of the sources in the field equations. One should take note of this subtlety.

Here we use the formulation of MAG, in which gravity is described by the set of fundamental field variables which consists of the independent metric $g_{ij}$ and connection $\Gamma_{ki}{}^j$. Such an approach was developed in \cite{Hehl:1976a,Hehl:1976b,Hehl:1976c,Hehl:1978,Hehl:1981,Obukhov:1982,Vassiliev:2005,Sotiriou:2007,Sotiriou:2010,Vitagliano:2011}, and this is alternative to the formalism which includes also a coframe field \cite{Hehl:1995} as a gravitational field variable. It is instructive to compare the field equations in the different formulations of MAG, and in particular, it is necessary to clarify the role and place of the {\it canonical} energy-momentum tensor as a source of the gravitational field. Since one does not have the coframe (tetrad) among the fundamental variables, the corresponding field equation is absent. Here we demonstrate that one can always rearrange the field equations of MAG in such a way that the canonical energy-momentum tensor is recovered as one of the sources of the gravitational field. 

For a large class of MAG models, the total Lagrangian of interacting gravitational and matter fields reads
\begin{equation}\label{Ltot}
{\mathfrak L} = {\mathfrak V} + F\,{\mathfrak L}_{\rm mat}.
\end{equation}
In general, the gravitational Lagrangian is constructed as a diffeomorphism covariant density function of the curvature, torsion, and nonmetricity, 
\begin{equation}\label{V}
{\mathfrak V} = {\mathfrak V}(g_{ij}, R_{ijk}{}^l, T_{ki}{}^j, Q_{kij}),
\end{equation}
whereas the matter Lagrangian depends on the matter fields $\psi^A$ and their {\it covariant derivatives} (\ref{Dpsi}):
\begin{equation}\label{Lmat}
{\mathfrak L}_{\rm mat} = {\mathfrak L}_{\rm mat}(g_{ij}, \psi^A, \nabla_i\psi^A).
\end{equation}

In the current literature a considerable attention is paid to the study of so-called modified models in which one assumes a possibility that the gravitational field may interact with matter in a nonminimal way. Accordingly, we here also allow for nonminimal interaction of the matter to the gravitational field via the {\it coupling function} 
\begin{equation}\label{F}
F = F(g_{ij}, R_{ijk}{}^l, T_{ki}{}^j, Q_{kij}).
\end{equation}

When $F = 1$, we recover the minimal coupling case. Let us write down the field equations of metric-affine gravity for that case. This can be done in several equivalent ways. The standard form is the set of the so-called ``first'' and ``second'' field equations. Using the covariant derivative for densities defined by (\ref{dA}), 
the field equations are given by
\begin{eqnarray}
\widehat{\nabla}{}_n{\mathfrak H}^{in}{}_k + {\frac 12}T_{mn}{}^i{\mathfrak H}^{mn}{}_k - {\mathfrak E}_k{}^i &=& - {\mathfrak T}_k{}^i,\label{1st}\\
\widehat{\nabla}{}_l{\mathfrak H}^{kli}{}_j + {\frac 12}T_{mn}{}^k{\mathfrak H}^{mni}{}_j - {\mathfrak E}^{ki}{}_j &=& {\mathfrak S}^i{}_j{}^k.\label{2nd}  
\end{eqnarray}
Here the generalized gravitational field momenta densities are introduced by
\begin{eqnarray}
{\mathfrak H}^{kli}{}_j &:=& -\,2{\frac {\partial {\mathfrak V}}{\partial R_{kli}{}^j}},\label{HH1}\\
{\mathfrak H}^{ki}{}_j &:=& -\,{\frac {\partial {\mathfrak V}}{\partial T_{ki}{}^j}},\label{HH2}\\
{\mathfrak M}^{kij} &:=& -\,{\frac {\partial {\mathfrak V}}{\partial Q_{kij}}},\label{HH}
\end{eqnarray}
and the gravitational hypermomentum and the generalized energy-momentum densities are constructed as
\begin{eqnarray}\label{EN}
{\mathfrak E}^{ki}{}_j &=& -\,{\mathfrak H}^{ki}{}_j - {\mathfrak M}^{ki}{}_j,\\
{\mathfrak E}_k{}^i &=& \delta_k^i\,{\mathfrak V} + {\frac 12}Q_{kln} {\mathfrak M}^{iln}\nonumber\\
&& +\,T_{kl}{}^n {\mathfrak H}^{il}{}_n + R_{kln}{}^m {\mathfrak H}^{iln}{}_m.\label{Eg}
\end{eqnarray}
The sources of the gravitational field are the canonical energy-momentum tensor density and the canonical hypermomentum density of matter, respectively:
\begin{eqnarray}
{\mathfrak T}_k{}^i &:=& {\frac {\partial {\mathfrak L}_{\rm mat}}{\partial\nabla_i\psi^A}}\,\nabla_k\psi^A - \delta^i_k{\mathfrak L}_{\rm mat},\label{canD}\\
{\mathfrak S}^i{}_j{}^k &:=& {\frac {\partial {\mathfrak L}_{\rm mat}}{\partial \Gamma_{ki}{}^j}} =  - {\frac {\partial {\mathfrak L}_{\rm mat}}{\partial\nabla_k\psi^A}} \,(\sigma^A{}_B)_j{}^i \psi^B.\label{tD}
\end{eqnarray}
The usual spin density arises as the antisymmetric part of the hypermomentum,
\begin{equation}
\tau_{ij}{}^k := {\mathfrak S}_{[ij]}{}^k,\label{spindef}
\end{equation}
whereas the trace ${\mathfrak S}^k = {\mathfrak S}^i{}_i{}^k$ is the dilation current density. The symmetric traceless part describes the proper hypermomentum \cite{Hehl:1995}.

It is straightforward to verify that instead of the first field equation (\ref{1st}), one can use the so-called zeroth field equation which reads
\begin{equation}
2\,{\frac {\delta {\mathfrak V}}{\delta g_{ij}}} = {\mathfrak t}^{ij}.\label{0th}
\end{equation}
On the right-hand side, the matter source is now represented by the metrical energy-momentum tensor which is defined by
\begin{equation}\label{tmet}
{\mathfrak t}_{ij} := 2\,{\frac {\partial {\mathfrak L}_{\rm mat}}{\partial g^{ij}}}.
\end{equation}
The system (\ref{1st}) and (\ref{2nd}) is completely equivalent to the system (\ref{0th}) and (\ref{2nd}), and it is a matter of convenience which one to solve. 

\subsection{Conservation laws}

The dynamics of extended bodies in metric-affine spaces can be derived by integrating the conservation laws. The latter are obtained from the Noether identities. After some algebra, the identities (\ref{Noetherall}) are recast into the following set of conservation laws 
\begin{eqnarray}
F{\mathfrak T}_k{}^i &=& F{\mathfrak t}_k{}^i + \widehat{\nabla}{}_n\left(F{\mathfrak S}^i{}_k{}^n\right),\label{cons1b}\\
\widehat{\nabla}{}_i\left(F{\mathfrak T}_k{}^i\right) &=& F \left({\mathfrak T}_l{}^i T_{ki}{}^l - {\mathfrak S}^m{}_n{}^l R_{klm}{}^n  - {\frac 12}{\mathfrak t}^{ij}Q_{kij} \right) \nonumber \\
&& -\,{\mathfrak L}_{\rm mat}\nabla_kF.\label{cons2b}
\end{eqnarray}
These relations hold ``on-shell'' when the matter variables $\psi^A$ satisfy the classical field equations, and are valid for general nonminimal coupling. 

\subsection{Rewriting the conservation laws}\label{nonmin_sec}

Using (\ref{dstar}) and decomposing the connection into the Riemannian and non-Riemannian parts, c.f. Eq. (\ref{dist}), we can recast the conservation law (\ref{cons1b}) into an equivalent form:
\begin{equation}
\check{\nabla}_j(F{\mathfrak S}^i{}_k{}^j) =  F({\mathfrak T}_k{}^i - {\mathfrak t}_k{}^i + N_{nm}{}^i{\mathfrak S}^m{}_k{}^n - N_{nk}{}^m{\mathfrak S}^i{}_m{}^n).\label{cons1c}
\end{equation}
In a similar way we can rewrite the conservation law (\ref{cons2b}). At first, with the help of (\ref{TN}) and (\ref{QN}) we notice that
\begin{equation}
F \left({\mathfrak T}_l{}^i T_{ki}{}^l - {\frac 12}{\mathfrak t}^{ij}Q_{kij}\right) = F({\mathfrak t}_l{}^i - {\mathfrak T}_l{}^i)N_{ki}{}^l + F{\mathfrak T}_l{}^i N_{ik}{}^l.\label{TTN}
\end{equation}
Then substituting here (\ref{cons1c}) and making use of (\ref{dstar}), (\ref{dist}), and the curvature decomposition (\ref{RRN}), after some algebra we recast (\ref{cons2b}) into
\begin{eqnarray}
\check{\nabla}_j\{&&F({\mathfrak T}_k{}^j + {\mathfrak S}^m{}_n{}^jN_{km}{}^n)\}\nonumber\\
&&= -\,F{\mathfrak S}^m{}_n{}^i(\widetilde{R}_{kim}{}^n - \widetilde{\nabla}_kN_{im}{}^n)\nonumber\\
&&\quad -\,{\mathfrak L}_{\rm mat}\widetilde{\nabla}_kF.\label{cons2c}
\end{eqnarray}
For the minimal coupling case, such a conservation law was derived in \cite{Neeman:1997,Obukhov:2006}. The importance of this form of the energy-momentum conservation law lies in the clear separation of the Riemannian and non-Riemannian geometrical variables. As we see, the post-Riemannian geometry enters (\ref{cons2c}) only in the form of the distortion tensor $N_{kj}{}^i$ which is coupled only to the hypermomentum current density ${\mathfrak S}^m{}_n{}^i$. This means that, in the {\it minimal coupling} case, ordinary matter -- i.e.\ without microstructure, ${\mathfrak S}^m{}_n{}^i=0$ -- does {\it not} couple to the non-Riemannian geometry. In contrast, in the {\it nonminimal coupling} case, the derivative of the coupling function $F$ on the right-hand side of (\ref{cons2c}) may lead to a coupling between non-Riemannian structures and ordinary matter.    

\subsection{Conserved current induced by a spacetime symmetry}\label{Induced_subsec}

As we have shown, every vector field generates a conserved current. It is straightforward to show that for a Lagrangian density of the form $\mathfrak{L} = F \mathfrak{L}_\mathrm{mat}$, as defined in \eqref{Lmat} and \eqref{F}, if we consider a generalized Killing vector field $\zeta^k$, the current \eqref{J1} reads, 
\begin{equation}
{\mathfrak J}^i := F\left[\zeta^k{\mathfrak T}_k{}^i - (\overline{\nabla}_m\zeta^n){\mathfrak S}^m{}_n{}^i\right].\label{I1}
\end{equation}
Here we have used the definitions \eqref{canD} and \eqref{tD}, and the conditions \eqref{LgMAG} and \eqref{LGMAG}. 

It is instructive to derive this result directly from the field equations. Namely, let us contract equation (\ref{cons1c}) with $\widetilde{\nabla}_i\zeta^k$ and contract equation (\ref{cons2c}) with $\zeta^k$, and then subtract the resulting expressions. Note that the contraction ${\mathfrak t}_k{}^i\widetilde{\nabla}_i\zeta^k = 0$ vanishes because the first factor is a symmetric tensor and the second one is skew-symmetric. Then after some algebra we find
\begin{equation}
\check{\nabla}_i{\mathfrak J}^i = F{\mathfrak S}^m{}_n{}^i{\cal L}_\zeta N_{im}{}^n - {\mathfrak L}_{\rm mat}{\cal L}_\zeta F.\label{DI1}
\end{equation} 
The right-hand side of (\ref{DI1}) depends linearly on the Lie derivatives along the Killing vector: ${\cal L}_\zeta F$ and ${\cal L}_\zeta N_{im}{}^n$, see (\ref{Ld}) and (\ref{LdC}). 

When $\zeta^k$ is a {\it generalized Killing} vector, we have ${\cal L}_\zeta N_{im}{}^n = 0$ in view of (\ref{LieN}). Furthermore, recalling that $F = F(g_{ij}, R_{klj}{}^i, T_{ki}{}^j, Q_{kij})$, we find 
\begin{eqnarray}
{\cal L}_\zeta F &=& {\frac {\partial F}{\partial g_{ij}}}{\cal L}_\zeta g_{ij} + {\frac {\partial F}{\partial R_{klj}{}^i}}{\cal L}_\zeta R_{klj}{}^i\nonumber\\
&& +\,{\frac {\partial F}{\partial T_{kj}{}^i}}{\cal L}_\zeta T_{kj}{}^i + {\frac {\partial F}{\partial Q_{kij}}}{\cal L}_\zeta Q_{kij} = 0,\label{LieF}
\end{eqnarray}
by making use of (\ref{Lg}), (\ref{LRMAG}), (\ref{LTMAG}), and (\ref{LQMAG}).

As a result, the right-hand side of (\ref{DI1}) vanishes for the generalized Killing vector field, and we conclude that the induced current (\ref{I1}) is conserved.

This generalizes the earlier results reported in \cite{Trautman:1973,Obukhov:2006,Obukhov:2007}. In Sec.~\ref{cons_mom} we will show that there is a conserved quantity constructed from the multipole moments which is a direct counterpart of the induced current (\ref{I1}). It is worthwhile to give an equivalent form of the latter:
\begin{equation}
{\mathfrak J}^i = F\left[\zeta^k({\mathfrak T}_k{}^i + {\mathfrak S}^m{}_n{}^iN_{km}{}^n) - (\widetilde{\nabla}_m\zeta^n){\mathfrak S}^m{}_n{}^i\right].\label{I2}
\end{equation}

\subsection{Test body equations of motion}

The details of the multipole approximation scheme are given elsewhere; see Appendix \ref{moments} for the definitions of the integrated moments. Here we limit ourselves to the pole-dipole equations of motion. 
Following \cite{Puetzfeld:Obukhov:2014:2}, we introduce the total orbital and the total spin angular moments
\begin{equation}
L^{ab} := 2p^{[ab]},\qquad S^{ab} := -2h^{[ab]},\label{LS}
\end{equation}
where $p^{ab}$ and $h^{ab}$ are generalized multipolar moments. The generalized total energy-momentum 4-vector and the generalized total angular momentum are given by
\begin{eqnarray}
{\cal P}^a &:=& F(p^a + N^a{}_{cd}h^{cd}) + p^{ba}\widetilde{\nabla}_bF,\label{Ptot}\\
{\cal J}^{ab} &:=& F(L^{ab} + S^{ab}).\label{Jtot}
\end{eqnarray}
Then the pole-dipole equations of motion take the form
\begin{eqnarray}
{\frac {D{\cal P}^a}{ds}} &=& {\frac 12}\widetilde{R}^a{}_{bcd}v^b{\cal J}^{cd} + Fq^{cbd}\widetilde{\nabla}^aN_{dcb} \nonumber\\ 
&& -\, \xi\widetilde{\nabla}^aF - \xi^b\widetilde{\nabla}_b\widetilde{\nabla}^aF,\label{DPtot}\\
{\frac {D{\cal J}^{ab}}{ds}} &=& -\,2v^{[a}{\cal P}^{b]} + 2F(q^{cd[a}N^{b]}{}_{cd} + q^{c[a|d|}N_{dc}{}^{b]}\nonumber\\ 
&& + q^{[a|cd|}N_d{}^{b]}{}_c) - 2\xi^{[a}\widetilde{\nabla}^{b]}F.\label{DJtot}
\end{eqnarray}
Here $v^a$ is the 4-vector of velocity of the test particle,  ${\frac {D}{ds}} = v^a\widetilde{\nabla}_a$, with $s$ the proper time of the particle, whereas $q^{abc}$, $\xi$ and $\xi^a$ are integrated moments relevant for the description. See App.\ \ref{moments}, as well as \cite{Puetzfeld:Obukhov:2014:2} for further details. 

\subsection{Conserved quantity for extended test bodies} \label{cons_mom}

The equations of motion for the multipole moments are derived from the conservation laws of the energy-momentum and the hypermomentum currents ${\mathfrak T}_k{}^i$ and ${\mathfrak S}^m{}_n{}^i$. In Sec.~\ref{Induced_subsec} we have demonstrated that every generalized Killing vector induces a conserved current depending on these two quantities, see Eqs. \eqref{I1} and \eqref{I2}. Quite remarkably, there is a direct counterpart of such an induced current built from the multipole moments. 

Let $\zeta^k$ be a generalized Killing vector, and let us contract equation (\ref{DPtot}) with $\zeta_a$ and equation (\ref{DJtot}) with ${\frac 12}\widetilde{\nabla}_a\zeta_b$, and then take the sum. This yields
\begin{eqnarray}
{\frac {D}{ds}}\left({\cal P}^a\zeta_a + {\frac 12}{\cal J}^{ab}\widetilde{\nabla}_a\zeta_b\right) &=& Fq^{cbd}{\cal L}_\zeta N_{dcb}\nonumber\\
&-& \xi\,{\cal L}_\zeta F - \xi^a\widetilde{\nabla}_a{\cal L}_\zeta F.\label{Dlie}
\end{eqnarray}
On the right-hand side of \eqref{Dlie} the Lie derivatives of the distortion tensor and of the coupling function vanish in view of (\ref{LieN}) and (\ref{LieF}). 

Consequently, we conclude that for every generalized Killing vector field the quantity
\begin{equation}\label{consPJ}
{\cal P}^a\zeta_a + {\frac 12}{\cal J}^{ab}\widetilde{\nabla}_a\zeta_b = {\rm const.}
\end{equation}
is conserved along a trajectory of an extended body. 

We thus observe a complete consistency between  (\ref{I1}) and (\ref{consPJ}), as well as between (\ref{DI1}) and (\ref{Dlie}).  

The conserved quantities with the similar structure in (\ref{consPJ}) were derived previously for extended test bodies with mass and spin in Einstein's general relativity \cite{Kuenzle:1972,Hojman:1975,Ehlers:Rudolph:1977} and in Einstein-Cartan gravity \cite{Trautman:1973}. Their various applications were recently discussed in \cite{Obukhov:Puetzfeld:2011,Hackmann:2014}. 

\subsection{Relation between theories of fields and particles} \label{mom_mom}

The similarity between the currents in a field-theoretic picture and the integrals of motion in particle mechanics is not occasional. Here we demonstrate that the current (\ref{I1}), (\ref{I2}) actually generates the conserved quantity (\ref{consPJ}) in the dipole approximation. Moreover, we will show that there exists a generalization of (\ref{consPJ}) to an arbitrary multipole order. 

As it is well established, the equations of motion of extended test bodies in external gravitational and other classical fields are derived from the corresponding conservations laws, see \cite{Mathisson:1937,Papapetrou:1951,Dixon:1964,Dixon:1967} for example. In simple terms, the dynamical description of an extended body is achieved by assigning to a body a set (infinite, in general) of multipole moments. The latter are determined by certain integrals of the conserved currents (energy-momentum tensor and other Noether currents) over the body. The multipole moments represent such characteristics of the body as its total mass, charge, momentum etc. By integrating the partial differential conservation laws, one obtains a system of ordinary differential equations of motion for the moments.

Here we apply the covariant multipole expansion of \cite{Dixon:1964} to the current (\ref{I1}). The method is based on Synge's \cite{Synge:1960} notion of the ``world function'' $\sigma$, which introduces a covariant generalization of the {\it finite} distance between the spacetime points $x$ and $y$. Basic definitions and notation are summarized in Appendix~\ref{moments}. By construction, this object $\sigma(x,y)$ depends on two arguments, and as a result the Synge formalism deals with arbitrary bitensor densities ${\mathfrak B}^{x_1 y_1}={\mathfrak B}^{x_1 y_1}(x,y)$. The most important technical tool is represented by the lemma \cite{Dixon:1964}:
\begin{eqnarray}
{\frac{D}{ds}} \int\limits_{\Sigma(s)}{\mathfrak B}^{x_1 y_1} d \Sigma_{x_1} &=& \int\limits_{\Sigma(s)} \widetilde{\nabla}_{x_1}{\mathfrak B}^{x_1 y_1} w^{x_2} d \Sigma_{x_2}\nonumber\\
&& + \,\int\limits_{\Sigma(s)} v^{y_2} \widetilde{\nabla}_{y_2} {\mathfrak B}^{x_1 y_1} d \Sigma_{x_1}.\label{int_aux}
\end{eqnarray}
We introduce the 4-velocity $v^{y_1}:=dx^{y_1}/ds$, with the proper time $s$, and denote ${\frac{D}{ds}} = v^i\widetilde{\nabla}_i$; the integrals are performed with an arbitrary bitensor density ${\mathfrak B}^{x_1 y_1}(x,y)$ over an arbitrary spatial hypersurface $\Sigma$. See \cite{Dixon:1964} for more details on the integrals and the construction of $w^x$. As in all our previous papers, we use a condensed notation suppressing the tensor indices so that $y_{n}$ denotes indices at the spacetime point $y$, etc. 

Following the standard procedure, we define the multipole moments for the current (\ref{I1}):
\begin{eqnarray}
j^{y_1\dots y_n} &:=& (-1)^n\!\!\int\limits_{\Sigma(s)}\!\!\sigma^{y_1}\cdots\sigma^{y_n}{\mathfrak J}^{x'}d\Sigma_{x'},\label{jmom}\\
i^{y_1\dots y_{n} y_0} &:=& (-1)^n\!\!\int\limits_{\Sigma(s)}\!\!\sigma^{y_1}\cdots \sigma^{y_n} g^{y_0}{}_{x_0}{\mathfrak J}^{x_0}w^{x'}d\Sigma_{x'}.\nonumber \\ \label{imom}
\end{eqnarray}
Then, by applying the lemma (\ref{int_aux}), we integrate the balance equation (\ref{DI1}). The result reads
\begin{eqnarray}
&&{\frac{D}{ds}}j^{a_1\dots a_n} = -\,nv^{(a_1}j^{a_2\dots a_n)} + n\,i^{(a_1\dots a_n)} \nonumber\\
&& + \,\bm{Q}^{a_1\dots a_{n}b}{}_c{}^d(F{\cal L}_\zeta N_{db}{}^c) - \bm{\Xi}^{a_1\dots a_n}({\cal L}_\zeta F).\label{djn}
\end{eqnarray} 
By boldface symbols we denote the following differential-algebraic operators 
\begin{eqnarray}
\bm{Q}^{a_1\dots a_{n}b}{}_c{}^d &=& \sum\limits_{k=0}^\infty\,{\frac {1}{k!}}\,q^{a_1\dots a_{n+k}b}{}_c{}^d\,\widetilde{\nabla}_{a_{n+1}}\cdots \widetilde{\nabla}_{a_{n+k}},\nonumber\\ && \label{bQ}\\
\bm{\Xi}^{a_1\dots a_{n}} &=& \sum\limits_{k=0}^\infty\,{\frac {1}{k!}}\,\xi^{a_1\dots a_{n+k}}\,\widetilde{\nabla}_{a_{n+1}}\cdots \widetilde{\nabla}_{a_{n+k}},\label{bX}
\end{eqnarray}
which are defined in terms of the multipole moments (\ref{Qmom}) and (\ref{Ximom}).

The current (\ref{I1}) has a nontrivial structure in that it is built from the Noether currents related to the diffeomorphism symmetry of the gravitational theory and of the auxiliary objects: vector fields $\zeta$ and the coupling function $F$. Taking this into account, we can use the lemma (\ref{int_aux}) once again to recast the moments into
\begin{eqnarray}
j^{a_1\dots a_n} &=& \bm{P}^{a_1\dots a_{n}b}(F\zeta_b) - \bm{H}^{a_1\dots a_{n}b}{}_c(F\overline{\nabla}_b\zeta^c),\label{jmom2}\\
i^{a_1\dots a_na_0} &=& \bm{K}^{a_1\dots a_{n}ba_0}(F\zeta_b) - \bm{Q}^{a_1\dots a_{n}b}{}_c{}^{a_0}(F\overline{\nabla}_b\zeta^c).\nonumber\\ && \label{imom2}
\end{eqnarray}
Here we introduced new operators
\begin{eqnarray}
\bm{P}^{a_1\dots a_{n}b} &=& \sum\limits_{k=0}^\infty\,{\frac {1}{k!}}\,p^{a_1\dots a_{n+k}b}\widetilde{\nabla}_{a_{n+1}}\!\cdots\! \widetilde{\nabla}_{a_{n+k}},\label{bP}\\
\bm{H}^{a_1\dots a_{n}b}{}_c &=& \sum\limits_{k=0}^\infty\,{\frac {1}{k!}}\,h^{a_1\dots a_{n+k}b}{}_c\widetilde{\nabla}_{a_{n+1}}\!\cdots\! \widetilde{\nabla}_{a_{n+k}},\label{bH}\\
 \bm{K}^{a_1\dots a_{n}bc} &=& \sum\limits_{k=0}^\infty\,{\frac {1}{k!}}\,k^{a_1\dots a_{n+k}bc}\widetilde{\nabla}_{a_{n+1}}\!\cdots\! \widetilde{\nabla}_{a_{n+k}},\label{bK}
\end{eqnarray}
defined in terms of the multipole moments of the hypermomentum and the energy-momentum currents (\ref{Tmom})-(\ref{Smom}). 

Let us now analyze the moments equation (\ref{djn}). This is an infinite system of ordinary differential equations, which is common for the equations of motion of extended test bodies. However, one equation is a special one in this system. It corresponds to $n=0$ and reads explicitly
\begin{eqnarray}
{\frac{d\,j}{ds}} = \bm{Q}^{b}{}_c{}^d(F{\cal L}_\zeta N_{db}{}^c) - \bm{\Xi}({\cal L}_\zeta F).\label{dj0}
\end{eqnarray}
We immediately observe its similarity to (\ref{DI1}). Moreover, the structure 
\begin{equation}
j = \bm{P}^{b}(F\zeta_b) - \bm{H}^{b}{}_c(F\overline{\nabla}_b\zeta^c)\label{jmom0}
\end{equation}
is obviously induced by the structure of the current (\ref{I1}). 

After these general derivations, we are now in a position to specialize to the pole-dipole case. 
In the dipole approximation, only the first terms proportional to the multipole moments up to the dipole order -- i.e.\ $p^a$, $p^{ab}$, $h^{ab}$, $k^{ab}$, $k^{abc}$, $q^{abc}$, $\xi$ and $\xi^a$ -- need to be taken into account in (\ref{djn})-(\ref{jmom0}). We then find for the equation (\ref{dj0}) and for the moment (\ref{jmom0}): 
\begin{eqnarray}
{\frac{d\,j}{ds}} &=& q^b{}_c{}^d\,F{\cal L}_\zeta N_{db}{}^c - \xi\,{\cal L}_\zeta F - \xi^a\widetilde{\nabla}_a{\cal L}_\zeta F,\label{dj_di}\\
j &=& p^a\,F\zeta_a + p^{ba}\widetilde{\nabla}_b(F\zeta_a) - h^b{}_a\,F\overline{\nabla}_b\zeta^a.\label{jmom_di}
\end{eqnarray}
Making use of (\ref{LS})-(\ref{Jtot}), we immediately see that these equations reproduce (\ref{Dlie}) and (\ref{consPJ}), respectively. 

\section{Conclusions}\label{conclusions_sec}

In this paper we have demonstrated that conserved currents can be naturally associated with spacetime diffeomorphisms (represented by vector fields on the spacetime manifold). Our findings extend previous results from Einstein's gravity with minimal coupling to the generalized metric-affine gravity theory with a possible nonminimal coupling. 

Mathematically, an important role in this construction is played by the generalized Killing vector fields, which represent the symmetries on the metric-affine spacetime. The corresponding formalism is based on the fundamental geometrical notions of the transposed connection and of the Lie derivatives which we heavily used in our previous studies \cite{Obukhov:2006,Obukhov:2007,Obukhov:2008}. 

There are many physically interesting applications of the results obtained. Among them is the possibility of the rigorous computation of the total mass and angular momentum for the exact solutions in the gravitational theories with and without torsion and nonmetricity. Another application is to use the conserved quantity (\ref{consPJ}) to simplify the study of the dynamics of extended microstructured test bodies in the generalized gravitational field models. Both issues are important for gravitational experiments, including space missions, which aim for an advanced probe of the geometrical structure of spacetime. 

The results of the present work should be used in the context of multipolar approximation schemes, which were recently worked out in \cite{Puetzfeld:Obukhov:2014:2,Obukhov:Puetzfeld:2015:1} for a very large class of gravitational theories, to further study the dynamics of test bodies. Of particular importance is the analysis of the detectability of post-Riemannian properties of spacetime, i.e.\ the torsion and the nonmetricity. It is worthwhile to mention that the recent literature on the Gravity Probe B experiment \cite{Will:2011} encompasses the misleading claim that the GPB result may set limits on torsion \cite{Hayashi:1990,Mao:2007}. These erroneous statements were corrected in \cite{Flanagan:2007,Puetzfeld:2007,Puetzfeld:2008,Hehl:2013pla,Puetzfeld:Obukhov:2014:2}, demonstrating that in minimally coupled gravitational theories the post-Riemannian geometry can only be detected with the help of microstructured matter. 

One may notice that the conditions of the generalized Killing vectors \eqref{LgMAG} and \eqref{LGMAG} imposed on the spacetime geometry may be quite strong. This is similar to the situation for the manifolds that admit various types of collineations. However, even in absence of exact symmetries one can consider approximate spacetime symmetries along the lines of \cite{Matzner:1968,Harte:2008}. We leave the corresponding analysis for the future. 

\section*{Acknowledgements}

This work was supported by the Deutsche Forschungsgemeinschaft (DFG) through the grant SFB 1128/1 (D.P.), by Comisi\'on Nacional de Investigaci\'on Cient\'ifica y Tecnol\'ogica (Chile) through grant CONICYT-PCHA/Mag\'isterNacional/2014-22141453 (F.P-O.) and Vicerector\'ia de Investigaci\'on y Desarrollo UdeC (VRID) through grant 214.011.058-1.0 (G.R.).

\appendix

\section{Notations and conventions}\label{notation}

In order to be consistent with our previous publications, we choose our main notations and conventions as those of \cite{Hehl:1995}. Most importantly, we stick to the definitions of \cite{Hehl:1995} for all the basic geometrical quantities such as the curvature, torsion, and nonmetricity, and we use the Latin alphabet to label the spacetime coordinate indices. 

The spacetime is modeled as a four-dimensional smooth manifold, and its metric has the signature $(+,-,-,-)$. It should be noted though that our definition of the metrical energy-momentum tensor differs by a sign from the definition used in \cite{Puetzfeld:Obukhov:2013}. 

In this work we are widely using tensor densities; we denote the densities by the ``Fraktur'' font to distinguish them from the tensor objects. 

\begin{table}
\caption{\label{tab_symbols}Directory of symbols.}
\begin{ruledtabular}
\begin{tabular}{ll}
Symbol & Explanation\\
\hline
\multicolumn{2}{l}{{Geometrical quantities}}\\
\hline
$g_{a b}$ & Metric\\
$\sqrt{-g}$ & Determinant of the metric \\
$x^{a}$, $s$ & Coordinates, proper time\\
$\zeta^a$ & Killing vector field\\
$\Gamma_{a b}{}^c$, $\overline{\Gamma}_{a b}{}^c$ & Connection, transposed conn. \\
$N_{a b}{}^c$ & Distortion \\
$Q_{a b c}$ & Nonmetricity \\
$T_{a b}{}^c$ & Torsion \\
$R_{a b c}{}^d$& Curvature \\
$R_{a b}$, $R$& Ricci tensor, scalar \\
$(\sigma^A{}_B)_i{}^j$ & Generators coord.\ transf.\\
$\delta^a_b$ & Kronecker symbol \\
${\mathfrak V}$ & Gravitational Lagrangian\\
${\mathfrak E}^{ki}{}_j$ & Gravitational hypermomentum \\
${\mathfrak E}_k{}^i$ & Gener.\ grav.\ energy-momentum \\
$\sigma$, $g^{y_0}{}_{x_0}$ & World function, parallel propagator\\
\hline
\multicolumn{2}{l}{{Matter quantities}}\\
\hline
$T^{a b}$ & Symmetric energy-momentum tensor\\
$\psi^A$ & General matter field \\
${\mathfrak L}_{\rm mat}$ & Matter Lagrangian \\
${\mathfrak J}^a$ & Generalized current\\
${\mathfrak K}^{ij}$ & Superpotential \\
${\mathfrak T}_k{}^i $ & Canonical energy-momentum \\
${\mathfrak S}^i{}_j{}^k $ & Canonical hypermomentum \\
$\tau_{ij}{}^k$ & Spin density\\
${\mathfrak t}^{ij}$ & Metrical energy-momentum \\
\hline
\multicolumn{2}{l}{{Auxiliary quantities}}\\
\hline
$\epsilon$ & Infinitesimal parameter \\
$\varepsilon^i$ & Arbitrary vector field \\
$\Phi^J$ & Multiplet of fields \\
$I$, ${\mathfrak L}$ & General action, Lagrangian \\
$F$ & Coupling function\\
${\mathfrak H}^{kli}{}_j$, ${\mathfrak H}^{ki}{}_j$ ,${\mathfrak M}^{kij} $ & Gener. gravitational field momenta\\
$p^{\dots}$, $k^{\dots}$, $h^{\dots}$, $q^{\dots}$, $\xi^{\dots}$,   & Test body integrated / generalized \\
$L^{ab}$, $S^{ab}$, ${\cal P}^a$, ${\cal J}^{ab}$, $v^a$ & moments, velocity \\
$\Omega^{\dots}$, $w^a$, $\Phi^{\dots}$, $\Psi^{\dots}$  & Auxiliary variables \\
\hline
\multicolumn{2}{l}{{Operators}}\\
\hline
${\cal L}_\zeta$ & Lie derivative\\
$\delta$ & Transformation under diffeomorph.\\ 
$\subsvar$ & Substantial variation \\
$\partial_i$, ${\nabla}_i$ & Partial, covariant derivative\\
${\stackrel * \nabla}{}_i$ & Modified cov.\ derivative \\
$\widehat{\nabla}_i$, $\check{\nabla}_i$ & Cov.\ density derivative, Riemannian \\ 
``$\widetilde{\phantom{AA}}$'' & Riemannian quantity \\
``$\overline{\phantom{AA}}$'' & Transposed quantity \\
``${\mathfrak A^{\dots}, \mathfrak B^{\dots}, \dots}$'' & Densities ``Fraktur''\\
$\bm{\Xi}^{\dots}$, $\bm{H}^{\dots}$, $\bm{K}^{\dots}$, & Auxiliary diff.\ operators \\
$\bm{P}^{\dots}$, $\bm{Q}^{\dots}$ & \\
\end{tabular}
\end{ruledtabular}
\end{table}

Table \ref{tab_symbols} displays the list of symbols used in the current paper. 

\section{Diffeomorphism invariance}\label{A1}

The explicit structure of the functions $\Omega_k{}^{i_1\cdots i_n}$, with $n=0,1,2,3$, is as follows
\begin{eqnarray}
\Omega_k &=& \partial_i\left({\frac {\partial {\mathfrak L}}{\partial \partial_ig_{mn}}}\partial_kg_{mn} + {\frac {\partial {\mathfrak L}}{\partial\partial_i\psi^A}} \,\partial_k\psi^A - \delta^i_k{\mathfrak L}\right)\nonumber\\
&& + {\frac {\delta {\mathfrak L}}{\delta g_{ij}}}\,\partial_kg_{ij} + {\frac {\delta {\mathfrak L}}{\delta\psi^A}}\,\partial_k\psi^A\nonumber\\
&& + {\frac {\partial {\mathfrak L}}{\partial \Gamma_{ln}{}^m}}\partial_k\Gamma_{ln}{}^m + {\frac {\partial {\mathfrak L}}{\partial \partial_i\Gamma_{ln}{}^m}}\partial_k\partial_i\Gamma_{ln}{}^m,\label{Om1}\\
\Omega_k{}^i &=& 2{\frac {\delta {\mathfrak L}}{\delta g_{ij}}}\,g_{kj} + {\frac {\delta {\mathfrak L}}{\delta\psi^A}}\,(\sigma^A{}_B)_k{}^i\,\psi^B\nonumber\\
&& + {\frac {\partial {\mathfrak L}}{\partial \partial_ig_{mn}}}\partial_kg_{mn} + {\frac {\partial {\mathfrak L}}{\partial\partial_i\psi^A}}\partial_k\psi^A  - \delta^i_k{\mathfrak L}\nonumber\\
&& + \partial_j\left(2{\frac {\partial {\mathfrak L}}{\partial \partial_jg_{in}}}g_{nk} + {\frac {\partial {\mathfrak L}}{\partial\partial_j\psi^A}}(\sigma^A{}_B)_k{}^i\psi^B\right)  \nonumber\\
&& + \,{\frac {\partial {\mathfrak L}}{\partial \Gamma_{li}{}^j}}\,\Gamma_{lk}{}^j + {\frac {\partial {\mathfrak L}}{\partial \Gamma_{il}{}^j}}\,\Gamma_{kl}{}^j - {\frac {\partial {\mathfrak L}}{\partial \Gamma_{lj}{}^k}}\,\Gamma_{lj}{}^i\nonumber\\
&& + \,{\frac {\partial {\mathfrak L}}{\partial \partial_i\Gamma_{ln}{}^m}}\,\partial_k \Gamma_{ln}{}^m   + {\frac {\partial {\mathfrak L}}{\partial \partial_n\Gamma_{il}{}^m}} \,\partial_n\Gamma_{kl}{}^m\nonumber\\
&& + {\frac {\partial {\mathfrak L}}{\partial \partial_n\Gamma_{li}{}^m}}\,\partial_n \Gamma_{lk}{}^m - {\frac {\partial {\mathfrak L}}{\partial \partial_n\Gamma_{lm}{}^k}}\,\partial_n\Gamma_{lm}{}^i ,\label{Om2}\\
\Omega_k{}^{ij} &=& {\frac {4\partial {\mathfrak L}}{\partial \partial_{(i}g_{j)n}}}g_{kn} + {\frac {2\partial {\mathfrak L}}{\partial\partial_{(i}\psi^A}} (\sigma^A{}_B)_k{}^{j)}\psi^B\nonumber\\
&& + {\frac {2\partial {\mathfrak L}}{\partial \Gamma_{(ij)}{}^k}} + {\frac {2\partial {\mathfrak L}}{\partial \partial_{(i}\Gamma_{j)l}{}^m}}\Gamma_{kl}{}^m \nonumber\\
&& + {\frac {2\partial {\mathfrak L}}{\partial \partial_{(i}\Gamma_{|l|j)}{}^m}}\,\Gamma_{lk}{}^m - {\frac {2\partial {\mathfrak L}}{\partial \partial_{(i}\Gamma_{|ln|}{}^k}}\,\Gamma_{ln}{}^{j)}, \label{Om3}\\
\Omega_k{}^{ijn} &=& {\frac {\partial {\mathfrak L}}{\partial \partial_{n}\Gamma_{(ij)}{}^k}} + {\frac {\partial {\mathfrak L}}{\partial \partial_{i}\Gamma_{(jn)}{}^k}} + {\frac {\partial {\mathfrak L}}{\partial \partial_{j}\Gamma_{(ni)}{}^k}}.\label{Om4}
\end{eqnarray}

\section{Multipole moments}\label{moments}

From the energy-momentum tensor density and the hypermomentum density, the integrated multipole moments of arbitrary order, $n = 0,1,2,\dots$, are defined by 
\begin{eqnarray}
p^{y_1\dots y_n y_0} &:=& (-1)^n\!\!\int\limits_{\Sigma(s)}\!\!\Phi^{y_1\dots y_n y_0}{}_{x_0}{\mathfrak T}^{x_0 x_1}d\Sigma_{x_1},\label{Tmom}\\
k^{y_2\dots y_{n+1} y_0 y_1} &:=& (-1)^n\!\!\int\limits_{\Sigma(s)}\!\!\Psi^{{y_2}\dots{y_{n+1} y_0 y_1}}{}_{x_0 x_1}\times\nonumber\\ 
&& \times{\mathfrak T}^{x_0 x_1}w^{x_2}d\Sigma_{x_2},\label{Kmom}\\
h^{y_2\dots y_{n+1}y_0 y_1} &:=& (-1)^n\!\!\int\limits_{\Sigma(s)}\Psi^{y_2 \dots y_{n+1}y_0 y_1}{}_{x_0 x_1 }\times\nonumber\\
&& \times{\mathfrak S}^{x_0 x_1 x_2}d\Sigma_{x_2},\label{Smom}\\
q^{y_3\dots y_{n+2}y_0 y_1 y_2} &:=& (-1)^n\!\!\int\limits_{\Sigma(s)}\!\!\Psi^{y_3 \dots y_{n+2} y_0 y_1}{}_{x_0 x_1} g^{y_2}{}_{x_2} \times\nonumber\\ 
&& \times{\mathfrak S}^{x_0 x_1 x_2 }w^{x_3}d\Sigma_{x_3},\label{Qmom}\\
\xi^{y_1\dots y_{n}} &:=& (-1)^n\!\!\int\limits_{\Sigma(s)}\!\!\sigma^{y_1}\cdots\sigma^{y_{n}}{\mathfrak L}_{\rm mat}w^{x_2}d\Sigma_{x_2}. \nonumber \\ \label{Ximom}
\end{eqnarray}
The integrals are taken over a cross-section $\Sigma(s)$ of the body's world tube. 
Here we introduced
\begin{eqnarray}
\Phi^{y_1\dots y_ny_0}{}_{x_0} &:=& \sigma^{y_1} \cdots \sigma^{y_n} g^{y_0}{}_{x_0},\label{Phi}\\
\Psi^{y_1\dots y_ny_0y'}{}_{x_0x'} &:=& \sigma^{y_1} \cdots \sigma^{y_n} g^{y_0}{}_{x_0}g^{y'}{}_{x'}.
\label{Psi}
\end{eqnarray}
In the derivation of the equations of motion we made use of the bitensor formalism; see, e.g., \cite{Synge:1960,DeWitt:Brehme:1960,Poisson:etal:2011} for introductions and references. In particular, the world function is defined as an integral $\sigma(x,y) := \pm{\frac 12}\left( \int\limits_x^y ds \right)^2$ over the geodesic curve connecting the spacetime points $x$ and $y$, where the upper/lower sign is chosen for timelike/spacelike curves, respectively. Note that our curvature conventions differ from those in \cite{Synge:1960,Poisson:etal:2011}. Indices attached to the world function always denote covariant derivatives, at the given point, i.e.\ $\sigma_y:= \nabla_y \sigma$; hence, we do not make explicit use of the semicolon in the case of the world function. The parallel propagator by $g^y{}_x(x,y)$ allows for the parallel transportation of objects along the unique geodesic that links the points $x$ and $y$. For example, given a vector $V^x$ at $x$, the corresponding vector at $y$ is obtained by means of the parallel transport along the geodesic curve as $V^y = g^y{}_x(x,y)V^x$. For more details see, e.g., section 5 in \cite{Poisson:etal:2011}. A compact summary of useful formulas in the context of the bitensor formalism, as well as a review of the multipolar formalism employed here can also be found in \cite{Obukhov:Puetzfeld:2015:1}.

\bibliographystyle{unsrt}

\bibliography{magconskill}

\end{document}